\documentclass[referee]{aa} 
\usepackage{epsfig}
\def\beq{\begin{equation}}
\def\eeq{\end{equation}}
\begin{document}

\title{Acceleration and Cyclotron Radiation, Induced by
Gravitational Waves}

\author{Demetrios Papadopoulos \inst{1}}
\institute{Department of Physics, Aristoteleion University of
Thessaloniki, \\54006 Thessaloniki, Greece \\
{\it email: papadop@astro.auth.gr} }

\offprints{D. Papadopoulos}

\date{Received/Accepted .. }

\titlerunning{ Acceleration and Cyclotron Radiation.. }
\authorrunning{D.B. Papadopoulos }

\abstract { The equations which determine the response of a
charged particle moving in a magnetic field to an incident
gravitational wave(GW) are derived in the linearized approximation
to general relativity. We briefly discuss several astrophysical
applications of the derived formulae taking into account the
resonance between the wave and the particle's motion which occurs
at $\omega_g=2\Omega$, whenever the GW is parallel to the
constant magnetic field. In the case where the GW is perpendicular
to the constant magnetic field, magnetic resonances appear at
$\omega_g=\Omega$ and $\omega_g=2\Omega$. Such resonant mechanism
may be useful to build models of GW driven cyclotron emitters.
}

\maketitle

\section{Introduction}

Papadopoulos and Esposito (1981) discussed the perturbations of
the Larmor orbits in the presence of a gravitational wave(GW) and
estimated the consequent magnetic bremsstrahlung. They has shown
that it is possible to identify the presence of GW in macroscopic
systems by detecting the shifts in the spectrum of the
electromagnetic radiation (cyclotron) given off by charged
particles as they interact with a GW.

Recently, the motion of a relativistic charged particle in a
constant magnetic field perturbed by GW incident along the
direction of the magnetic field has been examined(Jan-Willem van
Holten 1999 and references therein). In the same work, a
generalized energy conservation law to compute the variations of
the kinetic energy of the particle during the passage of the GW
has been derived and explicit computations in the orbit of the
charged particle due to the GW has been obtained.

In this paper, we discuss the interaction of GW with a charged
gyrating particle in the presence of a constant magnetic field
across the z-axis in the frame of linearized theory of gravity.

When the GW propagates parallel to the magnetic field with a
frequency $\omega_g$, the coupling of a gyrating particle with the
GW becomes very strong at the resonance which occurs between the
GW and the Larmor orbits. The resonance is at twice the Larmor
frequency($\Omega=\frac{eH}{mc}$) e.g $\omega_g=2\Omega$ (Macedo
and Nelson (1990)).

In the case that the GW propagates perpendicular to the magnetic
field, the interaction again becomes extremely efficient at the
resonances, $\omega=\Omega$ and $\omega=2\Omega$.

In both cases we verify that close to gyro resonances, the
obtained spectrum of the produced cyclotron radiation, becomes
comparable to the spectrum of the initially gyrating particle
especially in the vicinity of a source producing the GW .

Our results suggest that a) the linear theory breaks down at the
resonances and the interaction of the GW with gyrating particles
becomes very efficient and, b) even in the linear theory, support
the discrepancy on the estimations for the cyclotron damping
radiation recently discussed  by M. Servin (Martin Servin et al
(2001)) and Kleidis (K. Kleidis et al (1996); K. Kleidis et al
(1995)), since in Kleidis work, the problem is examined in the
non-linear theory where magnetic resonaces occur and some of them
are overlapped.

The paper organized as following. In sec.II, we derive the
equations of motion in the linearized theory. In sec.II we
discuss the interaction of the gravitational wave to the magnetic
field when the GW is parallel to the magnetic field. In sec.IV we
discuss the same problem assuming that the GW is perpendicular to
the magnetic field. The obtained results are discussed in sec.V

\newpage

\section{ Derivation of the equations of motion}

In the linearized approximation to general relativity the metric
tensor is decomposed in the fashion
\begin{eqnarray}
g_{ij}=\eta_{ij}+h_{ij} \end{eqnarray}

where the elements $h_{ij}$ are small compared to unity. By
imposing the condition

\begin{eqnarray} (h_{i}^{j}-\delta_{i}^{j} h_{l}^{l})_{;j}=0\end{eqnarray}

we reduce the vacuum field equations to homogeneous wave
equations for all components of $h_{i}^{j}$. The gravitational
field is then described by a symmetric traceless, divergenceless
tensor with two independent space components. Thus, the square of
the line element is
\begin{eqnarray}
ds^2=(\eta_{ij}+h_{ij})dx^{i}
dx^{j}=(dx^0)^2-(dx^a)^2+h_{\alpha\beta}dx^{\alpha}dx^{\beta}\end{eqnarray}

where Greek indices take values 1,2,3 and Latin 0,1,2,3.

The components of the covariant four-velocity, consistent with
the linearized theory, are
\begin{eqnarray}
u^0\equiv\frac{d x^0}{ds}\cong u_{(M)}^0[1-\frac{1}{2} h_{\alpha
\beta}u_{(M)}^{\alpha}u_{(M)}^{\beta}]\end{eqnarray}

\begin{eqnarray}
u^{\alpha}\equiv\frac{dx^{\alpha}}{dx^0}=\frac{dx^{\alpha}}{dx^0}\frac{d
x^0}{ds}\cong
u_{(M)}^0\frac{\upsilon^{\alpha}}{c}[1-\frac{1}{2}h_{\alpha\beta}u_{(M)}^{\alpha}u_{(M)}^{\beta}]\end{eqnarray}

where $u^0$, $u^{\alpha}$ are the components of the four-velocity
and the same quantities with the subscript M distinguish the
special-relativistic Minkowski values.

The equations of motion of a test particle, which is moving in
the presence of an electromagnetic field in the space-time
defined by Eq.(1) are given by

\begin{eqnarray} \frac{du^i}{ds}+\Gamma_{jk}^i u^{j}
u^{k}=\frac{e}{mc^2}F^{ik}u_{k}\end{eqnarray}

where the right-hand side is the inhomogeneous driving term
determined by the electromagnetic field in the space-time defined
by the Eq.(1)

For the metric (1), the nonzero Christoffel symbols are

\begin{eqnarray}
\Gamma_{\alpha\beta}^0=-\frac{1}{2}h_{\alpha\beta,0},~~\Gamma_{0\beta}^{\alpha}=\frac{1}{2}h_{\beta,0}^{\alpha},\nonumber\\
\Gamma_{\beta\gamma}^{\alpha}=\frac{1}{2}(h_{\beta,\gamma}^{\alpha}+h_{\gamma,\beta}^{\alpha}-h_{\beta\gamma}^{\alpha})\end{eqnarray}

From Eqs.(6),(4),(5) and (7), the equations of motions take the
form:

\begin{eqnarray} u_{(M)}^0 u_{(M),0}^0
[1-h_{\alpha\beta}u_{(M)}^{\alpha}u_{(M)}^{\beta}]-\frac{1}{2}(u_{(M)}^0)^2[h_{\alpha\beta}u_{(M)}^{\alpha}u_{(M)}^{\beta}]_{,0}-\frac{1}{2}h_{\alpha\beta,0}u_{(M)}^{\alpha}u_{(M)}^{\beta}\nonumber\\
=\frac{e}{mc^2}(\eta^{0 l}-h^{0
l}-\frac{\eta^{0l}}{2}h_{\alpha\beta}u_{(M)}^{\alpha}u_{(M)}^{\beta})(F_{l
0} u^{0}+F_{l \alpha} u^{\alpha})\end{eqnarray}

and
\begin{eqnarray}
u_{(M)}^0 u_{(M),0}^{\alpha}
[1-h_{\alpha\beta}u_{(M)}^{\alpha}u_{(M)}^{\beta}]-\frac{1}{2}u_{(M)}^0
u_{(M)}^{\alpha}
[h_{\alpha\beta}u_{(M)}^{\alpha}u_{(M)}^{\beta}]_{,0}\nonumber\\
+h_{\beta,0}^{\alpha}u_{(M)}^0u_{(M)}^{\beta}+\frac{1}{2}(h_{\beta,\gamma}^{\alpha}+h_{\gamma,\beta}^{\alpha}-h_{\beta\gamma}^{\alpha})u_{(M)}^{\beta}u_{(M)}^{\alpha}\nonumber\\
=\frac{e}{mc^2}(\eta^{\alpha l}-\frac{\eta^{\alpha
l}}{2}h_{\alpha\beta}u_{(M)}^{\alpha}u_{(M)}^{\beta}-h^{\alpha l
})(F_{l 0}u_{(M)}^0+F_{l\alpha}u_{(M)}^{\alpha})\end{eqnarray}

where
\begin{eqnarray}
u_{(M)}^0=(1-\frac{\upsilon^2}{c^2})^{-1/2},~~and~~u_{(M)}^{\alpha}=\frac{\upsilon^{\alpha}}{c}(1-\frac{\upsilon^2}{c^2})^{-1/2}
\end{eqnarray}

Finally, the equations of motion (8) and (9), in the Newtonian
and linearized limit, reduce to the equations:
\begin{eqnarray}
\frac{\partial{\upsilon^i}}{{\partial t}}+\eta^{ik}
h_{jk,t}\upsilon^j+\frac{1}{2}\eta^{il}[h_{jl,k}+h_{kl,j}-h_{jk,l}]\upsilon^j
\upsilon^k=\frac{q}{mc}(\eta^{ia}-h^{ia})F_{al}\upsilon^l\end{eqnarray}

To make further progress with the derived equations of motion
(11), we consider the gravitational wave which is characterized
by the wave vector
\begin{eqnarray}
k^{\alpha}=\frac{\omega}{c}(\sin(\theta),0,\cos(\theta)),~~and~~,\frac{\omega^2}{c^2}=k^a
k_{a}\end{eqnarray} and one of two possible states of
polarization given by

\begin{eqnarray} h_{ij}=h_0 (e_{i}^1 e_{j}^1-e_{i}^2
e_{j}^1)\exp{[i\frac{\omega_g}{c}(x^1\sin(\theta)+x^3\cos(\theta)-ct)]}\end{eqnarray}

where $h_0$ is the amplitude of the gravitational wave and
$\omega_g=2\pi\nu_g$ is the angular frequency of the GW.

The vectors $\mathbf{e}^1$ and $\mathbf{e}^2$ have space
components only and satisfy the conditions

\begin{eqnarray} e^{1\mu}e_{\mu}^1=e^{2\mu}e_{\mu}^2=1,~~ with,~~ k^{\mu}
e_{\mu}^1=k^{\mu} e_{\mu}^2=0\end{eqnarray}

Conditions (14) imply
\begin{eqnarray}
\mathbf{e}_{\mu}^1=(\cos(\theta),0,-\sin(\theta)),~~\mathbf{e}_{\mu}^2=(0,1,0)\end{eqnarray}

Under the above consideration we proceed to the following two
cases, a) the GW is parallel to the magnetic field and b) the GW
is perpendicular to the magnetic field which will be analyzed in
the following two section.
\newpage

\section{ The GW is parallel to the magnetic field}

We choose the electromagnetic field to be

\begin{eqnarray} F_{ij}=\left (
\begin{array}{cccc}
0 & 0 & 0 & 0 \\
0 & 0 &-H_3 & 0 \\
0 & H_3 & 0 & 0\\
0 & 0 & 0 & 0
\end{array}
\right ) \end{eqnarray}

where $H^3=H=constant$.

We choose the gravitational wave to propagate parallel to the
magnetic field e.g., in Eq.(13) we obtain $\theta=0$ and thus
$h=h_{11}=-h_{22}=h_0\exp{(\frac{i\omega_g}{c}(z-ct))}$. Eq.(11)
with the aid of Eq.(16) yields:

\begin{eqnarray}\frac{\partial{\upsilon^1}}{\partial t}-\Omega
\upsilon^2=-hi\omega_g\upsilon^1[1-\frac{\upsilon^3}{c}]\end{eqnarray}

\begin{eqnarray}\frac{\partial{\upsilon^2}}{\partial t}+\Omega
\upsilon^1=hi\omega_g\upsilon^2[1-\frac{\upsilon^3}{c}]\end{eqnarray}

\begin{eqnarray}\frac{\partial{\upsilon^3}}{\partial t}=-ih\frac{1}{2c}\omega_g
[(\upsilon^1)^2-(\upsilon^2)^2]\end{eqnarray}

where $\Omega=\frac{eH}{mc}$.
 To solve the system of Eqs.(17-19), we decompose the
components of the 3-velocity as follows:

\begin{eqnarray}\upsilon^1\simeq \upsilon_0^1+\upsilon_1^1,~~\upsilon^2\simeq
\upsilon_0^2+\upsilon_1^2,~~\upsilon^3\simeq
0+\upsilon_1^3\end{eqnarray} where the subscript zero means zero
order in the sense that $h_0=0$, while the subscript one means
first order in the sense that $h_0\ne0$.

The perturbed equations of motion are derived from Eqs.(17-19) and
(20). Thus, after some straightforward calculations (see Appendix
B) we find the solution(Macedo and Nelson (1990)):

\begin{eqnarray} \upsilon^1&\simeq& \upsilon_{0T}\cos(\Omega t+a)+ h_0
\upsilon_{0T}\frac{\Omega-\omega_g}{(2\Omega-\omega_g)}
\{\cos{[k_g z+(\Omega-\omega_g)t]}-\cos{(k_g z-\Omega
t)}\}\end{eqnarray}
\begin{eqnarray}\upsilon^2&\simeq& -\upsilon_{0T}\sin(\Omega t+a)+h_0
\upsilon_{0T}\frac{\Omega-\omega_g}{(2\Omega-\omega_g)}
\{\sin{[k_g z+(\Omega-\omega_g)t]}-\sin{(k_g z-\Omega
t)}\}\end{eqnarray}
\begin{eqnarray}\upsilon^3&\simeq&
\frac{h_0}{2}(\frac{\upsilon_{0T}^2}{c})\exp{(ik_g
z)}\{\frac{\omega_g^2}{4\Omega^2-\omega_g^2}
-[\frac{\omega_g}{(2\Omega-\omega_g)}\exp{(i(2\Omega-\omega_g)t}\nonumber\\&-&\frac{\omega_g}{(2\Omega+\omega_g)}\exp{(i(2\Omega+\omega_g)t}]\}
\end{eqnarray}

where $\upsilon_{0T}^2=\upsilon_{x}^2+\upsilon_y^2=constant$ and
$a=constant$.

It is evident that in Eqs.(21-23) if $h_0=0$, we obtain the
components of the space velocity of the initial gyrating charged
particle. If $h_0\not=0$, the Eqs.(21-23) reveal that the
gyrating charged particle diverts from its initial plane orbit
and moves into a helical trajectory. Now the vector
$\vec{\upsilon}=(\upsilon^1,\upsilon^2,\upsilon^3)$ does not move
in a circle, but on the surface of cone with its axis along the
$\vec{H}$. From Eqs.(21-23) we conclude that the particle is
accelerated at the resonace $\omega_g=2\Omega$. The existence of
the resonace at $\omega_g=2\Omega$, in Eqs.(21-23) is due to GW.
Because of the resonance the gyrating particle gains kinetic
energy. Thus, if $E_0$ and $E_1$ are the kinetic energy of the
particle before and after the interaction with the GW,
respectively; the energy gained by the particle in a period, let
say T, is given by the average of the ration $\frac{E_1}{E_0}$
e.g.

\begin{eqnarray}
I=\frac{1}{T}\int_0^T\frac{E_1}{E_0}dt&\approx&1-2h_0\frac{\Omega-\omega_g}{2\Omega-\omega_g}\cos{(k_gz)}
\end{eqnarray}

Obviously, as $\omega_g$ approaches $2\Omega$ taking value
between $\Omega$ and $2\Omega$, the factor,
$\frac{\Omega-\omega_g}{2\Omega-\omega_g}$, becomes negative
making the term multiplied by $h_0$, to approach to plus infinity
with positive values. This suggests extra emission of cyclotron
radiation which will change the spectra distribution of the
radiation of the initial gyrating charged particle and transfer
of energy from the GW to the particle.

Integrating the Eqs.(21-23) we obtain the parametric equations of
motion of the charged gyrating particle interacting with a GW in
the presence of a constant magnetic field across the z-axis.
These are
\begin{eqnarray}
x_{(1)}(t)&=&x_{(01)}+h_0
\upsilon_{0T}\frac{\Omega-\omega_g}{2\Omega-\omega_g}\{ \frac
{\sin{[k_g
z+(\Omega-\omega_g)t]}}{\Omega-\omega_g}-\frac{\sin{(k_g z-\Omega
t)}}{\Omega} \}\end{eqnarray}
\begin{eqnarray} y_{(1)}(t)&=&y_{(01)}-h_0 \upsilon_{0T}\frac{\Omega-\omega_g}{2\Omega-\omega_g}\{ \frac {\cos{[k_g
z+(\Omega-\omega_g)t]}}{\Omega-\omega_g}+\frac{\cos{(k_g z-\Omega
t)}}{\Omega} \}\end{eqnarray} and
\begin{eqnarray}
z_{(1)}(t)&=&z_{(01)}+\frac{h_0}{2}\upsilon_{0T}(\frac{\upsilon_{0T}}{c})\exp{(ik_g
z)}\{\frac{t \omega_g^2}{4\Omega^2-\omega_g^2}\nonumber\\
&+&\frac{i}{2}[\frac{\omega_g}{(2\Omega-\omega_g)^2}\exp{(i(2\Omega-\omega_g)t}+\frac{\omega_g}{(2\Omega+\omega_g)^2}\exp{(i(2\Omega+\omega_g)t}]\}
\end{eqnarray}
where $x_{(01)}$,$y_{(01)}$ and $z_{(01)}$ are constants of
integration.

The intensity of the radiation per solid angle per unit interval
of frequency produced from a charge test particle moving in the
presence of a magnetic field which interacts with the GW, maybe
obtained from the relation(Landau 1975):
\begin{eqnarray}
\frac{ d^2 I}{d \Omega_a d\omega'}&=&\frac {q^2(\omega')^2}{4\pi^2
c} \vert \int_{-\infty}^{\infty} dt
\exp{(i\omega'(t-\frac{\bf{n}. \bf{R}}{c}))}
 [\bf{n}\times(\bf{n} \times \bf{b})]  \vert^2\end{eqnarray}
where $\omega'$ is the frequency of the outgoing radiation,
$\bf{n}=\sin{(\theta)}\bf{i}+\cos{(\theta)}\bf{k}$, $\bf{R}$ is a
vector which joins the charged particle with the observer,
$\bf{b}$ is the velocity of the charge particle.

We carry out the integral of Eq.(28) neglecting terms of the
order $\frac{\upsilon^2}{c^2}$. Thus, we find:
\begin{eqnarray} \frac{ d^2 I}{d \Omega_a d \omega^{'}}&=&\frac
{q^2\varpi^2}{c} \sum_{-\infty}^{\infty} \delta
(l\Omega-\varpi)\{[\cot^2{(\theta)}J_l^2(\Phi)+\frac{\upsilon_{0T}^2}{c^2}(J_l^{'})^2(\Phi)]\nonumber\\
&-&4h_0\cot^2{(\theta)}\cos{(k_g
z)}\frac{\Omega-\omega_g}{2\Omega-\omega_g} J_l^2(\Phi)\}
\end{eqnarray} where
$\varpi=l\Omega[1+h_0\frac{\upsilon_{0T}^2\omega^2}{2c^2(2\Omega^2-\omega_g^2)}\sin{(k_g
z)}]=l\Omega[1+O(1/c^2)]$,
$\Phi=\frac{\omega^{'}}{\Omega}\frac{\upsilon_{0T}}{c}\sin{(\theta)}$,
$J_{j}$ is the Bessel function of first kind and $J_{l}^{'}$ its
first ordinary derivative in terms of its argument. For
simplicity we call
\begin{eqnarray}
L&=&[\cot^2{(\theta)}J_l^2(\Phi)+\frac{\upsilon_{0T}^2}{c^2}(J_l^{'})^2(\Phi)]\end{eqnarray}
and \begin{eqnarray}T&=&-4h_0\cot^2{(\theta)}\cos{(k_g
z)}\frac{\Omega-\omega_g}{2\Omega-\omega_g}
J_l^2(\Phi)\end{eqnarray}

It is evident that, as $\omega_g$ approaches $2\Omega^{+}$, the
factor $\frac{\Omega-\omega_g}{2\Omega-\omega_g}\rightarrow
\infty$ making the term T tend to minus infinity. But, If
$\omega_g$ approaches $2\Omega$ taking value between $\Omega$ and
$2\Omega$, the factor,
$\frac{\Omega-\omega_g}{2\Omega-\omega_g}$, becomes negative
making the term T, to approach to plus infinity with positive
values. Also, from Eq.(31), we see that the divergence of the
term T is faster as we approach to the source producing GW.
Nevertheless, in the linearized theory of gravitation, we have to
approach the resonant in such a way that the term T remains always
below the term $L$, otherwise the linear theory breaks down.

\section{ The GW is perpendicular to the magnetic field}

We shall now consider the case where the GW propagates
perpendicular to the unperturbed magnetic field
$H^{\mu}=(0,0,H^3)=const.$ Thus, in a reference frame where
$H^{\mu}$ has the z-direction (as for the previous case), but the
GW propagates along the x-direction, the two non-vanishing
components of the GW are given by Eqs.(13) setting $\theta=\pi/2$
and
\begin{eqnarray} h_{33}=-h_{22}=h_0\exp{\frac{i \omega}{c}(x-ct)}\end{eqnarray}
Subsequently, from Eqs.(11),(20) and (32) we obtain the following
equations of motion:
\begin{eqnarray}
\upsilon^1&=&\upsilon_0^1+\upsilon_1^1=\upsilon_{0T}\cos{(\Omega
t+a)}+\upsilon_{0T}h_0 \{ C\cos{(\Omega t)}\nonumber\\
&-& [A\cos{(k_g z-\omega_g t)}-B\sin{(k_g z-\omega_g t)}]
\}\end{eqnarray}

\begin{eqnarray}
\upsilon^2&=&\upsilon_0^2+\upsilon_1^2=-\upsilon_{0T}\sin{(\Omega
t+a)}-\upsilon_{0T}h_0 \{C\sin{(\Omega t)}\nonumber\\
&+&[A\sin{(k_g z-\omega_g t)}-B\cos{(k_g z-\omega_g t)}]
\}\end{eqnarray} and
\begin{eqnarray}
\upsilon^3&=&\upsilon_0^3+\upsilon_1^3=\sigma-\sigma(1-\frac{\sigma}{c})h_{22}
\end{eqnarray}

where, $\sigma=constant$ and may be chosen equal zero,

\begin{eqnarray} C&=&\frac{\Omega^2}{(\Omega-\omega_g)(2\Omega-\omega_g)}[1
+(\frac{\upsilon_{0T}}{c})\frac{\omega_g^2(2\Omega-\omega_g)}{\Omega(\Omega+\omega_g)(3\Omega-\omega_g)}]\end{eqnarray}
\begin{eqnarray}
A&=&\frac{\Omega}{\Omega-\omega_g}-\frac{\Omega}{2\Omega-\omega_g}\cos{(\Omega
t+2a)}\nonumber\\
&+&\frac{\upsilon_{0T}}{4c}[\frac{\omega_g}{\Omega-\omega_g}-\frac{\omega_g(3\Omega+\omega_g)}{(\Omega+\omega_g)(3\Omega-\omega_g)}\cos{(2\Omega
t+2a)}]\end{eqnarray} and
\begin{eqnarray}B&=&\frac{\Omega-\omega_g}{2\Omega-\omega_g}\sin{(\Omega
t+2a)}-\frac{\upsilon_{0T}}{2c}\frac{\omega_g^2}{(\Omega+\omega_g)(3\Omega-\omega_g)}
\sin{(2\Omega t+2a)}\end{eqnarray}

Obviously, if in Eqs.(33-35)  $h_0=0$, then we obtain the space
velocities of the initial gyrating charged particle.

If $h_0 \not=0$, then we have gyrating motion again, but magnetic
resonances appear at $\omega_g=\pm \Omega$,$\omega_g=2\Omega$ and
$\omega_g=3\Omega$.

The existence of the above resonances, in Eqs.(33-35) is due to
GW. Because of those resonances the gyrating particle gains
kinetic energy. Thus, as in section (III) we verify that the
energy gained by the particle in a period, let say T, is given
averaging the ration $\frac{E_1}{E_0}$ e.g.

\begin{eqnarray}
I_2&=&\frac{1}{T}\int_0^T\frac{E_1}{E_0}dt\approx1+2h_0
C\approx1+2h_0
\frac{\Omega^2}{(\Omega-\omega_g)(2\Omega-\omega_g)}\end{eqnarray}

Obviously, if $\omega_g$ approaches $2\Omega$ from the right or
$\Omega$ from the left, the factor,
$\frac{\Omega^2}{(\Omega-\omega_g)(2\Omega-\omega_g)}$, becomes
positive making the term multiplied by $h_0$, to approach to plus
infinity with positive values suggesting changes to the spectra
distribution of the radiation of the initial gyrating charged
particle.

Integrating Eqs.(33-35) we derive the parametric equations of
motion which are:
\begin{eqnarray}
x(t)&=&\frac{\upsilon_{0T}}{c}\sin{(\Omega t)}+h_0\upsilon_{0T}
[\frac{\sin(\Omega t)}{\Omega} C+X_{h}]\end{eqnarray}

\begin{eqnarray} y(t)&=&\frac{\upsilon_{0T}}{c}\cos{(\Omega t)}+h_0\upsilon_{0T}
[\frac{\cos(\Omega t)}{\Omega} C+Y_{h}]\end{eqnarray}

\begin{eqnarray} z(t)&=&\sigma
t+i\frac{\sigma}{\omega_g}(1-\frac{\sigma}{c})h_{22}
\end{eqnarray}

where, $\sigma=constant$ and may be chosen equal zero, the
expression for  $X_{h}$ and $Y_{h}$ are given explicitly in the
Appendix A.

In the Eqs.(40) and (41) a drift term, with resonances at
$\omega_g=\Omega, \omega_g=2\Omega$, is present. This drift term
can generate electric currents. Those currents are sources of
secondary electromagnetic waves. For further details see (Macedo
and Nelson (1990)).

Following the same procedure as in in section III, Eq.(28) reduce
\begin{eqnarray} \frac{ d^2 I}{d \Omega_a d \omega^{'}}&=&\frac
{q^2\varpi^2}{c} \sum_{-\infty}^{\infty} \delta
(l\Omega-\varpi)\{[\cot^2{(\theta)}J_l^2(\Phi)+\frac{\upsilon_{0T}^2}{c^2}(J_l^{'})^2(\Phi)]\nonumber\\
&-&h_0\cos{(2\theta)}\cot^2{(\theta)}(\frac{\upsilon_{0T}}{c})\frac{2\Omega^2}{(\Omega-\omega_g)(2\Omega-\omega_g)}J_l^2(\Phi)\}
\end{eqnarray}

As in sec.III, we call
\begin{eqnarray}
L&=&[\cot^2{(\theta)}J_l^2(\Phi)+\frac{\upsilon_{0T}^2}{c^2}(J_l^{'})^2(\Phi)]\end{eqnarray}

and
\begin{eqnarray}
T_2&=&-h_0\cos{(2\theta)}\cot^2{(\theta)}(\frac{\upsilon_{0T}}{c})\frac{2\Omega^2}{(\Omega-\omega_g)(2\Omega-\omega_g)}J_l^2(\Phi)\end{eqnarray}

Now the term $T_2$ has two magnetic resonances. It is evident
that, as $\omega_g$ approaches $2\Omega^{-}$, or $\Omega^{+}$ and
$\theta\in(0,\frac{\pi}{4})$, the term $T_2$ tend to plus
infinity. Also, from Eq.(45), we see that the divergence of the
term $T_2$ becomes faster as we approach to the source producing
GW and as we are dealing with ultra relativistic particles where
ratio $\frac{\upsilon_{0T}}{c}$ takes higher values. Again, as in
in the section (III), in the linearized theory of gravitation, we
have to approach to the resonances carefully!, in the sense that
the term $T_2$ should not exceed term $L$, otherwise the theory
breaks down.

\newpage

\section{ Conclusions}

In this article we pose the following problem. If $h_0=0$, we
reproduce a well known formula for the spectrum distribution of a
gyrating charged particle in both cases e.g when the GW is
parallel and perpendicular to the magnetic field. In this case
the angular distribution of the gyro radiation is highly
anisotropic. The radiation is concentrated mainly in the plane of
the orbit.

If $h_0\not= 0$, we are dealing with the interaction of a GW with
a gyrating charged particle. We have distinguished two cases:

(a). The GW propagates parallel to the constant magnetic field:

Because of the GW, the gyrating charged particle diverts from its
initial plane orbit and starts to move across a helical
trajectory. Also, because of the GW, the spectrum of the produced
cyclotron radiation isolated by a factor proportional to $h_0$,
namely T, in which a resonant at $\omega_g=2\Omega$ appears. If
$\omega_g$ approaches $2\Omega$ taking values in the interval
$I_1=(\Omega,2\Omega)$,the term T, approaches to plus infinity
indicating that the existence of the resonant interaction between
the charged particle and the GW, can lead to a strong emission of
cyclotron radiation, even in the linear theory of gravity. The
suggested mechanism of cyclotron radiation could be useful to the
astrophysicists especially in the case they make simultaneously
observations of the so obtained cyclotron radiation and the GW.
Therefore, knowing that astrophysicists are looking to detect GW
at frequencies between $\nu_g=(10^2-10^3) Hz$ (Cutler, Thorne
2002), magnetic resonance may occur whenever the magnetic fields
could be between $H\approx(1.-2.)G$ ($\nu_g=10^2Hz$),
$H\approx(10-21)G$ ($\nu_g=10^3Hz$) and in both cases the T term
becomes positive and comparable to the L term . But, if we want
to use electromagnetic radiation for indirect detection of GW,
then other frequencies are also important, e.g. lower frequencies
of $\nu_g=(1-10) Hz$ coming from binary neutron stars before
coalescence, the corresponding magnetic fields are
$H\approx(0.01-0.02)G$ ($\nu_g=1Hz$), $H\approx(0.1-0.2)G$
($\nu_g=10Hz$)or high frequencies $\nu_g=(10^3-10^4) Hz$, due to
supernovae explosions, to normal modes of pulsating neutron stars
or stellar size black holes. In the latter case the corresponding
magnetic fields are $H\approx(107-214)G$.

Outside interval $I_1$ the particle seems to lose energy due to
destructive interference of the two oscillators e.g. the
sinusoidal gravitational wave and the gyrating particle. In this
case, the actual loss described by the negative term $T$ is of
the order $h_0$.

(b). The GW propagates in the x-direction e.g perpendicular to the
constant magnetic field.

In this case, we verify that, 1) the gyrating particle remains on
the initial plain of orbit described by Eqs.(33-35) and for a
certain $\theta=\theta_0$, let's say, $\theta_0\in(0,
\frac{\pi}{4})$, at $\omega_g=\Omega$ and $\omega_g=2\Omega$ the
corresponding term $T_2$ diverges.  2) If $\omega_g$ approaches
to $\Omega$ from the left, or to $2\Omega$ from the right, whereas
$\theta_0\in(\frac{\pi}{4}, \frac{\pi}{2})$, the term $T_2$
becomes positive, and approaches to plus infinity, unless
$\theta_0\in(0, \frac{\pi}{4})$. For the frequencies mentioned
above, two magnetic resonaces may occur for the same values of
$\omega_g$ at the same range of the magnetic field. But if
$\omega_g$ approaches to one or the other resonance taking values
in the interval $I_1$, whereas
$\theta_0\in(\frac{\pi}{4},\frac{\pi}{2})$, the term $T_2$ becomes
negative (unless $\theta_0\in(0, \frac{\pi}{4})$), indicating that
the particle lose energy due to the same reason mentioned in
paragraph (a). Another interesting feature is that the term $T_2$
is proportional to the ratio $\frac{\upsilon_{0T}}{c}$, which
means that high relativist particles support the term $T_2$ to
become more significant. Nevertheless, in both cases, the
strength of the cyclotron radiation described by the ratios
$\frac{T}{L_1}$ and $\frac{T_2}{L_1}$, remain constant for large
values of l(large frequencies). This may seen in figure 1 where
we plot the above mentioned ratios for $\theta_0=80^{0}$,
obtaining the amplitude of the GW to be $h_0=10^{-21}$ and its
frequency $\nu_g=10^3$.

\begin{figure}[h]
\centering\epsfig{file=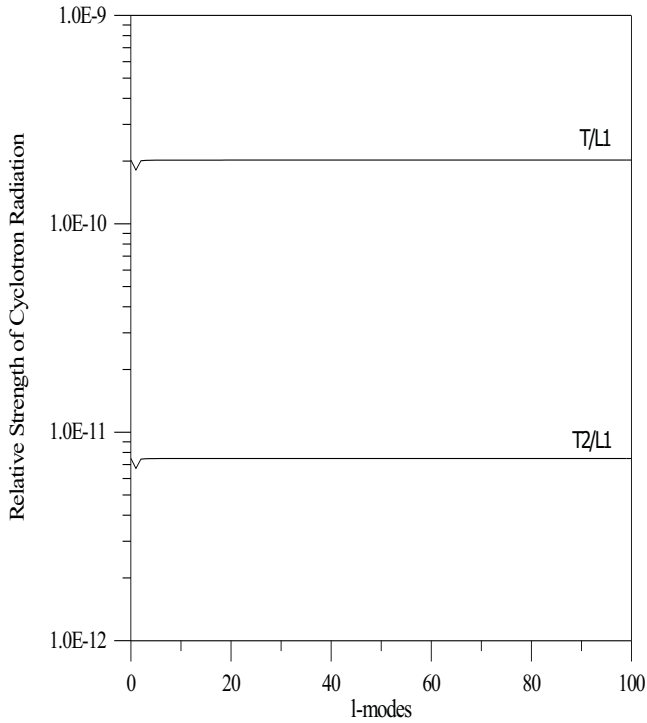,width=8.5cm,height=9.5cm}
\caption{The strengths of cyclotron radiation in the cases where
$H$ is parallel to the $k$(T/L1) and $H$ is perpendicular to the
$k$}
\end{figure}
However, we have to point out that, in both cases, the terms
multiplied by $h_0$ should not exceed the term $L$, otherwise the
linearized theory of gravity breaks down.

In the nonlinear theory, the problem could be more interesting.
The problem somehow has been examined from a dynamical point of
view, considering the equations of motion from a Hamiltonian, (
Varvoglis and Papadopoulos (1992), Kleidis Varvoglis and
Papadopoulos(1993), Kleidis Varvoglis, Papadopoulos and Esposito
(1995) and Kleidis Varvoglis and Papadopoulos(1996)), and
integrating them numerically. In this case, the interaction of
the GW with gyrating charged particle exhibits resonances and in
several cases chaotic behavior. We intend to discuss the problem
in the non linear theory in a forthcoming paper.

{\bf Acknowledgements:} The author would like to thank Loukas
Vlahos, Kostas Kokkotas L.Witten and Nik Stergioulas for their
comments, criticism and beneficial discussions.

\newpage
{\bf Appendix A}

In the Eqs.(33-35) the$X_h$ and $Y_h$ are:

\begin{eqnarray} X_h&=&-\{-\frac{\Omega}{\omega_g(\Omega-\omega_g)}\sin{(k_g
x-\omega_g
t)}-\frac{\Omega}{2(2\Omega-\omega_g)}[-\frac{1}{\Omega+\omega_g}\sin{[k_g
x-(\Omega+\omega_g)t]}\nonumber\\
&+&\frac{1}{\Omega-\omega_g}\sin{[k_g x +(\Omega-\omega_g)t]}]
+\frac{\Omega-\omega_g}{2(2\Omega-\omega_g)}[\frac{1}{\Omega+\omega_g}\sin{[k_g
x-(\Omega+\omega_g)t]}\nonumber\\
&+&\frac{1}{\Omega-\omega_g}\sin{[k_g x+(\Omega-\omega_g)t]}]
+\frac{\upsilon_{0T}}{4c}[-\frac{1}{(\Omega-\omega_g)}\sin{(k_g
x-\omega_g t)}\nonumber\\
&-&\frac{\omega_g(3\Omega+\omega_g)}{2(\Omega+\omega_g)(3\Omega-\omega_g)}[-\frac{1}{2\Omega+\omega_g}\sin{[k_g
x-(2\Omega+\omega_g)t]}+\frac{1}{2\Omega-\omega_g}\sin{[k_g x
+(2\Omega-\omega_g)t]}]\nonumber\\
&+&\frac{\omega_g^2}{2(\Omega+\omega_g)(3\Omega+\omega_g)}[\frac{1}{2\Omega+\omega_g}\sin{[k_g
x-(2\Omega+\omega_g)t]}+\frac{1}{2\Omega-\omega_g}\sin{[k_gx+(2\Omega-\omega_g)t]}]\nonumber\\
&-&\sin{k_g
x}\{\frac{\Omega}{\omega_g(\Omega-\omega_g)}-\frac{\Omega(\Omega-2\omega_g)}{(2\Omega-\omega_g)(\Omega^2-\omega_g^2)}+\frac{2\Omega\omega_g^2}{(\Omega+\omega_g)(3\Omega-\omega_g)(4\Omega^2-\omega_g^2)}\nonumber\\
&+&\frac{\upsilon_{0T}}{4c}[\frac{1}{\Omega-\omega_g}+\frac{\omega_g^2(3\Omega+\omega_g)}{(2\Omega+\omega_g)(3\Omega-\omega_g)(4\Omega^2-\omega_g^2)}+\frac{2\Omega\omega_g^2}{(\Omega+\omega_g)(3\Omega-\omega_g)(4\Omega^2-\omega_g^2)}]\}
\end{eqnarray} and
\begin{eqnarray}
Y_h&=&-\{-\frac{C}{\Omega}+\frac{\Omega}{\omega_g(\Omega-\omega_g)}\cos{(k_g
x-\omega_g t)}-\frac{\Omega}{2(2\Omega-\omega_g)}[\frac{\cos{(k_g
x-(\Omega-\omega_g)t)}}{\Omega-\omega_g}+\frac{\cos{(k_g
x-(\Omega+\omega_g)t)}}{\Omega+\omega_g}\nonumber\\
&+&\frac{\Omega-\omega_g}{2(2\Omega-\omega_g)}[\frac{\cos{(k_g
x-(\Omega-\omega_g)t)}}{\Omega-\omega_g}-\frac{\cos{(k_g
x-(\Omega+\omega_g)t)}}{\Omega+\omega_g}]\nonumber\\
&-&\cos{(k_g
x)}[\frac{\Omega}{\omega_g(\Omega-\omega_g)}-\frac{\Omega^2}{(2\Omega-\omega_g)(\Omega^2-\omega_g^2)}+\frac{\omega_g}{(2\Omega-\omega_g)(\Omega+\omega_g)}]\nonumber\\
&+&\frac{\upsilon_{0T}}{4c}[\frac{1}{(\Omega-\omega_g)}\cos{(k_g
x-\omega_g
t)}+\frac{2\omega_g}{(\Omega-\omega_g)(3\Omega-\omega_g)}sin{(k_g
x-\omega_g t)}\sin{(2\Omega t)}\nonumber\\
&-&\frac{\omega_g(3\Omega+\omega_g)}{2(\Omega+\omega_g)(3\Omega-\omega_g)}[\frac{\cos{(k_g
x-(2\Omega-\omega_g)t)}}{2\Omega-\omega_g}+\frac{\cos{(k_g
x-(2\Omega+\omega_g)t)}}{2\Omega+\omega_g}]\nonumber\\
&-&\cos{(k_g
x)}[\frac{1}{\Omega-\omega_g}-\frac{2\Omega\omega_g(3\Omega+\omega_g)}{(\Omega+\omega_g)(3\Omega-\omega_g)(4\Omega^2-\omega_g^2)}]\}
\end{eqnarray}
\newpage
{\bf Appendix B}

We start with the Eqs.(17-19)

\begin{eqnarray}\frac{\partial{\upsilon^1}}{\partial t}-\Omega
\upsilon^2&=&-hi\omega_g\upsilon^1[1-\frac{\upsilon^3}{c}]\end{eqnarray}

\begin{eqnarray}\frac{\partial{\upsilon^2}}{\partial t}+\Omega
\upsilon^1&=&hi\omega_g\upsilon^2[1-\frac{\upsilon^3}{c}]\end{eqnarray}

\begin{eqnarray}\frac{\partial{\upsilon^3}}{\partial t}&=&-ih\frac{1}{2c}\omega_g
[(\upsilon^1)^2-(\upsilon^2)^2]\end{eqnarray}

We write the Eqs.(48),(49) as follows:

\begin{eqnarray}\frac{\partial{(\upsilon^1+i\upsilon^2)}}{\partial
t}+\Omega[\upsilon^1+i
\upsilon^2]&=&hi\omega_g(\upsilon^1-i\upsilon^2)[1-\frac{\upsilon^3}{c}]\end{eqnarray}

In order to solve the Eq.(51), we decompose the components of the
3-velocity as follows:

\begin{eqnarray}\upsilon^1\simeq \upsilon_0^1+\upsilon_1^1,~~\upsilon^2\simeq
\upsilon_0^2+\upsilon_1^2,~~\upsilon^3\simeq
0+\upsilon_1^3\end{eqnarray} where the subscript zero means zero
order in the sense that $h_0=0$, while the subscript one means
first order in the sense that $h_0\ne0$.

We substitute Eqs.(52) into Eq.(51) and the perturbed equation now
reads :

\begin{eqnarray}\frac{\partial{[(\upsilon_0^1+\upsilon_1^1)+i(\upsilon_0^2+i\upsilon_1^2)]}}{\partial
t}\nonumber&+&i\Omega[(\upsilon_0^1+\upsilon_1^1)+i(\upsilon_0^2+\upsilon_1^2)]\nonumber\\
&=&hi\omega_g[(\upsilon_0^1+\upsilon_1^1)-i(\upsilon_0^2+\upsilon_1^2][1-\frac{\upsilon_1^3}{c}]\end{eqnarray}

{\bf I. Zero Order Equations}

\begin{eqnarray}\frac{\partial{(\upsilon_0^1+i\upsilon_0^2)}}{\partial
t}&=&-i\Omega(\upsilon_0^1+i\upsilon_0^2)\end{eqnarray}

This yields

\begin{eqnarray}\upsilon_0^1=\upsilon_{0T}\cos(\Omega t+a)
~~and~~\upsilon_0^2=-\upsilon_{0T}\sin(\Omega t+a)\end{eqnarray}

where $a=constan$ and
$\upsilon_{0T}^2=\upsilon_x^2+\upsilon_y^2=constan$.
\newpage
{\bf II. First Order equations}

From Eq.(53), we have

\begin{eqnarray}
\frac{\partial{(\upsilon_1^1+i\upsilon_1^2)}}{\partial
t}+i\Omega(\upsilon_1^1+i\upsilon_1^2)
&=&hi\omega_g(\upsilon_0^1-i\upsilon_0^2)\end{eqnarray}

Notice that on right hand side, the factor
$h(\upsilon_0^1-i\upsilon_0^2)$ gives

\begin{eqnarray}
h(\upsilon_0^1-i\upsilon_0^2)&=&h_0\upsilon_{0T}e^{i(k_g
z-\omega_g t)}[\cos{(\Omega t+a)}-i\sin{(\Omega
t+a)}]\nonumber\\&=&h_0\upsilon_{0T} e^{i(k_g z-\omega_g
t)}e^{i(\Omega t+a)}
\end{eqnarray}

Now from Eqs. (56) and (57) we have

\begin{eqnarray}
\frac{\partial{(\upsilon_1^1+i\upsilon_1^2)}}{\partial
t}+i\Omega(\upsilon_1^1+i\upsilon_1^2)&=&h_0\upsilon_{0T}
e^{i(k_g z+a)}e^{it(\Omega -\omega_g)}\end{eqnarray}

We treat Eq.(58) as an ordinary first order differential equation
with the initial conditions, if $t=0$ then
$(\upsilon_1^1+i\upsilon_1^2)=0$. Thus, we have

\begin{eqnarray}
(\upsilon_1^1+i\upsilon_1^2)&=&e^{-it\Omega}\{C+ih_0\upsilon_{0T}\omega_ge^{i(k_g
z+a)} \int e^{it(\Omega-\omega_g)}e^{i\int\Omega
 dt} dt\}\nonumber\\
&=&e^{-it\Omega}\{C+ih_0\upsilon_{0T}\omega_ge^{i(k_g z+a)} \int
e^{it(2\Omega-\omega_g)} dt\}\nonumber\\
&=&e^{-it\Omega}\{C+ih_0\upsilon_{0T}\frac{\omega_g}{2\Omega-\omega_g}e^{i(k_g
z+a)}e^{it(2\Omega-\omega_g)}\}
\end{eqnarray}

Upon the consideration of the initial conditions we have
\begin{eqnarray}
C&=&-h_0\upsilon_{0T}\frac{\omega_g}{2\Omega-\omega_g}e^{i(k_g
z+a)}\end{eqnarray}
Eventually, from the Eqs.(59) and (60) we
obtain

\begin{eqnarray}
(\upsilon_1^1+i\upsilon_1^2)&=&h_0\upsilon_{0T}\frac{\omega_g}{2\Omega-\omega_g}e^{i(k_g
z+a)}[e^{it(\Omega-\omega_g)}-e^{-it\Omega }]\end{eqnarray}

or

\begin{eqnarray}\upsilon_1^1&=&h_0 \upsilon_{0T}\frac{\omega_g}{2\Omega-\omega_g}\{\cos{[k_g
z+(\Omega-\omega_g)t]}-\cos{(k_g z-\Omega t)}\}\end{eqnarray}

\begin{eqnarray}
\upsilon_1^2&=&h_0
\upsilon_{0T}\frac{\omega_g}{2\Omega-\omega_g}\{\sin{[k_g
z+(\Omega-\omega_g)t]}-\sin{(k_g z-\Omega t)}\}\end{eqnarray}

Furthermore, following the same method, from Eq.(19)  we find
\begin{eqnarray}
\upsilon_1^3&=&\frac{h_0}{2}\upsilon_{0T}(\frac{\upsilon_{0T}}{c})\exp{(ik_g
z)}\{\frac{\omega_g^2}{4\Omega^2-\omega_g^2}\nonumber\\
&-&[\frac{\omega_g}{(2\Omega-\omega_g)}\exp{(i(2\Omega-\omega_g)t}-\frac{\omega_g}{(2\Omega+\omega_g)}\exp{(i(2\Omega+\omega_g)t}]\}
\end{eqnarray}

\end{document}